\documentclass[12pt,twoside]{article}

\usepackage{amssymb}
\usepackage{amsmath}
\usepackage{latexsym}
\usepackage{longtable}
\usepackage{epsfig}
\usepackage{subfig}
\usepackage{graphicx,bbm,psfrag}
\graphicspath{{images/}}

\begin{document}
\begin{titlepage}
\begin{center}
\vspace*{2cm} {\Large {\bf Radiative damping: a case
study\bigskip\bigskip\\}}
{\large{Herbert Spohn}}\bigskip\bigskip\\
   { Zentrum Mathematik and Physik Department, TU M\"unchen,\\
 D-85747 Garching, Germany\\e-mail:~{\tt spohn@ma.tum.de}}\\

\end{center}
\vspace{5cm} \textbf{Abstract.} We are interested in the motion of a
classical charge coupled to the Maxwell self-field and subject to a
uniform external magnetic field, $B_0$. This is a physically
relevant, but difficult dynamical problem, to which contributions
range over more than one hundred years. Specifically, we will study
the Sommerfeld-Page approximation which assumes an extended charge
distribution at small velocities. The memory equation is then linear
and many details become available. We discuss how the friction
equation arises in the limit of ``small'' $B_0$ and contrast this
result with the standard Taylor expansion resulting in a second
order equation for the velocity of the charge.

\end{titlepage}

\section{Introduction}
Radiative damping must be taken into account in the design of
synchrotron radiation sources \cite{2}. Amongst other tasks, ultra
strong lasers are expected to provide a quantitative test of
radiative damping forces in the ultra relativistic regime \cite{1}.
For stellar jets and other highly accelerated beams of charged
particles radiative friction plays a dominant role \cite{2}. The ground 
work on radiative friction was accomplished already 
Abraham \cite{Ab} and Lorentz \cite{Lo}, who undertook a vast effort to
derive effective equations of motion for the charged particle
starting from a model of an extended charged body coupled to the
Maxwell field. Later on fully relativistic models followed. I refer
to recent studies \cite{Ki,Wa} and to the monographs \cite{Ro, Ya,Sp}, in
which prior work is put in perspective. To be clear, we will consider
classical charges coupled to the classical electromagnetic field.
When the interest is in the radiative decay of, say, an excited
hydrogen atom, then a quantum version of the theory is required,
which is beyond the scope of this 
note.

Granted a few exceptions, in all derivations of the effective equations 
of motion for the charged particle one uses Taylor expansion in a small parameter, which leads to 
a differential equation of second order in the velocity. In such 
a procedure it is never detailed how to match the true initial data with 
the initial conditions for the effective dynamics.
In addition the second order differential equation generates spurious 
solutions which are a mere artifact of the approximation method. The case study 
presented here is so simple that both defects can be easily exhibited and remedied.
While we discuss the extension to more physical models, we believe that a satisfactory 
derivation of the equations of radiative friction are still to be accomplished.

For the sake of discussion I prefer to fix a very concrete physical set-up,
namely the motion of a charged particle subject to a uniform
external magnetic field and coupled to its self-generated Maxwell
field.
 The coupling to the self-field makes our dynamical problem
difficult. To keep the presentation as simple as possible and to be
able to provide a complete argument, we will study an oversimplified
model known as Sommerfeld-Page equation.  In the
Sommerfeld-Page equation the charge distribution of the particle is
taken to be uniform over a sphere of radius $R$ and normalized to
$e$. The charge distribution is supposed to be rigid and independent
of velocity in the given inertial frame. Also small velocities are
assumed. At the initial time $t=0$ one has to specify position and
velocity of the center of the charge distribution and the
self-field. One natural choice for the self-field is the Coulomb
field generated by the charge distribution at rest. An even more
physical choice for the self-field would be the field as if the
particle has been traveling with constant velocity $v_0$ for all
$t\leq 0$. There are other choices and the initial Maxwell
self-field is to some extent arbitrary. For sure, finite energy and some 
smoothness is required. On the other hand, only as one example, an incident plane wave  would
have to be taken into account explicitly in the Sommerfeld-Page
equation.

With these assumptions the equations for the motion of the particle
read
\begin{equation}\label{1.2}
m_0\frac{d}{dt}\boldsymbol{v}(t)=\frac{e B_0}{c}
\boldsymbol{v}(t)^\perp +\frac{e^2}{12\pi c R^2}
\big(\boldsymbol{v}(t-2c^{-1}R)-\boldsymbol{v}(t)\big)\,,
\end{equation}
\begin{equation}\label{1.2a}
m_0\frac{d}{dt}v_3=\frac{e^2}{12\pi c R^2}
\big(v_3(t-2c^{-1}R)-v_3(t)\big)\,.
\end{equation}
Here $m_0$ is the mechanical (bare) mass of the particle, $e$ the
charge, and $c$ the velocity of light. The magnetic field points
along the 3-direction with magnitude $B_0$. The velocity is
decomposed in the component $\boldsymbol{v}$ orthogonal and the
component $v_3$ parallel to the magnetic field. If
$\boldsymbol{v}=(v_1,v_2)$, then $\boldsymbol{v}^\perp=(-v_2,v_1)$.
In the literature (\ref{1.2}), (\ref{1.2a}) is referred to as
Sommerfeld-Page equation \cite{So,Pa}, see also Appendix A for
more details on the approximations involved. We will discuss only
(\ref{1.2}), since (\ref{1.2a}) is automatically included as the
special case $B_0=0$ in (\ref{1.2}).

To streamline the notation, we set
\begin{equation}\label{1.3}
\omega=\frac{e B_0}{m_0 c}\,,\quad \tau=\frac{2 R}{c}\,,\quad
\alpha=\frac{e^2}{3\pi m_0 c^3 \tau^2}\,.
\end{equation}
Thus $\omega$ measures the strength of the external magnetic field
and $\alpha$ the coupling strength to the self-field. (\ref{1.2})
becomes then
\begin{equation}\label{1.4}
\frac{d}{dt}\boldsymbol{v}(t)=\omega\boldsymbol{v}(t)^\perp +\alpha
\big(\boldsymbol{v}(t-\tau)-\boldsymbol{v}(t)\big)\,.
\end{equation}
The natural dimensionless parameters will be $\omega\tau$ and
$\alpha\tau$. (\ref{1.4}) is a delay equation because the current
change in velocity depends on the past history. Also the notion
differential delay equation is very common. In
statistical mechanics one uses the term memory equation, indicating
again the dependence on the past. Eq. (\ref{1.4}) has to be
supplemented with the initial condition
\begin{equation}\label{1.5}
\boldsymbol{v}(t)=\boldsymbol{u}(t)\,,\quad 0\leq t\leq \tau\,,
\end{equation}
with some specified function $\boldsymbol{u}(t)$ encoding the
interaction with the self-field up to time $\tau$, an information
which usually is not available in explicit form.  In fact we will see that the precise
details of $\boldsymbol{u}(t)$, except for the average and the end
point $\boldsymbol{u}(\tau)$, play no role.

To obtain from (\ref{1.2}) an effective differential equation seems
to be an easy task. Since $R$ is small, one Taylor expands in $R$ up
to second order and arrives at
\begin{equation}\label{1.5a}
\big(m_0+\frac{e^2}{6\pi c^2 R}\big)\frac{d}{dt}\boldsymbol{w}(t)
=\frac{e B_0}{c}\boldsymbol{w}(t)^\perp +\frac{e^2}{6\pi
c^3}\frac{d^2}{dt^2}\boldsymbol{w}(t)\,,
\end{equation}
where we use $\boldsymbol{w}(t)$ to distinguish from the true
solution $\boldsymbol{v}(t)$. Formally one expects that
\begin{equation}\label{1.5b}
\boldsymbol{v}(t)-\boldsymbol{w}(t)=\mathcal{O}(R)\,.
\end{equation}
The prefactor of $\dot{\boldsymbol{w}}(t)$ is regarded as effective
mass,
\begin{equation}\label{1.5d}
m_\mathrm{eff}=m_0 +\frac{e^3}{3\pi c^3 \tau}=(1+\alpha \tau)m_0\,.
\end{equation}
If one really wants $R\to 0$, following Dirac \cite{Di}, one takes the
double limit $R\to 0$, $m_0\to -\infty$ with $m_\mathrm{eff}>0$
fixed. Then it is claimed that
\begin{equation}\label{1.5c}
\lim_{R\to 0} \boldsymbol{v}(t)=\boldsymbol{w}(t)\,.
\end{equation}

To fix a unique solution to (\ref{1.5a}) one has to prescribe
$\boldsymbol{w}(0)$ and $\dot{\boldsymbol{w}}(0)$. On the other hand
for (\ref{1.4}) we need to specify a whole initial function
$\boldsymbol{u}(t)$, $0\leq t\leq \tau$. A complete derivation of
(\ref{1.5a}), or (\ref{1.5c}), thus must give a prescription how to
link $\boldsymbol{u}(t)$ to $\boldsymbol{w}(0)$ and
$\dot{\boldsymbol{w}}(0)$. This part is missing 
and the Taylor expansion does not tell us.

In fact, the situation is even more complicated. The phase space for
(\ref{1.5a}) is $\mathbb{R}^2\times \mathbb{R}^2$. Embedded is the two-dimensional
plane $\mathcal{C}_\omega$ with $0\in\mathcal{C}_\omega$. If
$\big(\boldsymbol{w}(0),\dot{\boldsymbol{w}}(0)\big) \in
\mathcal{C}_\omega$, then the motion is stable and spirals towards
0. On the other hand for initial data off $\mathcal{C}_\omega$, the
motion diverges exponentially fast to infinity. Since such run away
solutions have never been observed, I regard them as unphysical. But
then one has to explain why for a broad class of initial
$\boldsymbol{u}(t)$, $0\leq t\leq \tau$, one always has initial data
for (\ref{1.5c}) which lie in $\mathcal{C}_\omega$. Again the Taylor
expansion does not tell us.

For the Sommerfeld-Page equation a straightforward alternative to
the Taylor expansion (\ref{1.5a}) is available. Eq. (\ref{1.4}) is the standard
and most elementary example of a linear differential-difference
equation which is exhaustively covered in the literature
\cite{BeCo,Mi}. I will use here the conventional method for
construction of solutions. Given the linear structure, it is rather
surprising that so far the Sommerfeld-Page equation has been analyzed only
rudimentarily \cite{LMS,MS,Ro1,Oc,Ro2,GTS}.

Our task is to study the solution to (\ref{1.4}), (\ref{1.5}).
Roughly one expects that the synchrotron motion is damped and the
particle will come to rest in the long time limit. In principle one
could explore properties of (\ref{1.4}) for all $\omega$ and
$\alpha$. But it is of advantage to keep the physical parameter
range in mind. The classical electron radius is defined by the
condition that the mechanical rest energy equals the electrostatic
energy of the charge distribution. This yields $R\simeq 10^{-13}$
cm. Then $\alpha\tau=10^{-5}$ and for a magnetic field of $10^6$
Tesla one arrives at $\omega\tau=10^{-2}$. The strongest laboratory
fields are of order 10 Tesla. Thus it will be safe to eventually
consider only the regime $0 < \alpha\tau\ll 1$, $|\omega\tau|\ll 1$.

Before entering into a more detailed analysis, let me summarize
already now the main findings.\medskip\\
(1)\textit{ Zero external magnetic field,} $B_0=0$.\smallskip\\
\textit{(i)} If $m_{\mathrm{eff}}/m_0>0$, then the motion is stable.
On the other hand if $m_{\mathrm{eff}}/m_0<0$, generically the
solution increases exponentially. Expressed in dependence on $m_0$,
if $m_0>0$, then $m_\mathrm{eff}>0$, and the motion is stable. If
$-e^2/3\pi c^3\tau^2<m_0<0$, then the motion is unstable but still
$m_\mathrm{eff}>0$. If $m_0<-e^2/3\pi c^3\tau^2$, then
$m_\mathrm{eff}<0$ and the motion is
again stable.\smallskip\\
\textit{(ii)} If $m_{\mathrm{eff}}/m_0>0$, then
$\boldsymbol{v}(t)\to\overline{\boldsymbol{v}}$ exponentially fast
as $t\to\infty$, where
\begin{equation}\label{1.7}
\overline{\boldsymbol{v}}=(1+\alpha\tau)^{-1}\big(\boldsymbol{u}(\tau)+\alpha\int^\tau_0
\textrm{d}s \boldsymbol{u}(s)\big)\,,
\end{equation}
to say, there is a decay constant  $\gamma_{\alpha\tau} < 1$ such that
\begin{equation}\label{1.8}
|\boldsymbol{v}(n\tau)-\overline{\boldsymbol{v}}|\leq c_0
(\gamma_{\alpha\tau})^n
\end{equation}
with integer $n>0$ and some constant $c_0$. At $ \alpha\tau =
0.278$, the decay constant $\gamma_{\alpha\tau} \approx 0.07$ and
decreases rapidly as $\alpha\tau$ decreases to $0$. At $ \alpha\tau
= -1$, the decay constant
$\gamma_{\alpha\tau} \approx 0.25$.\medskip\\
(2) \textit{Non-zero external magnetic field,} $B_0\neq 0$,
$|\omega\tau|\ll 1$, $m_0>0$.\smallskip\\
\textit{(iii)} If $m_0>0$, $|\omega\tau|\ll 1$, and $\omega\tau\neq
0$, then $\boldsymbol{v}(t)\to 0$ exponentially fast as $t\to\infty$
with an upper bound
\begin{equation}\label{1.9}
|\boldsymbol{v}(n\tau)|\leq
c_0\Big(1-\frac{1}{2}\frac{\alpha\tau}{1+\alpha\tau}
\big(\frac{\omega\tau}{1+\alpha\tau}\big)^2\Big)^n\,.
\end{equation}
\smallskip\\
\textit{(iv)} For $m_0>0$ and at fixed $R$, if $|\omega\tau|\ll 1$,
the solution to (\ref{1.4}) is well approximated by
\begin{equation}\label{1.10}
(1+\alpha\tau)\frac{d}{dt}\boldsymbol{w}(t)=\omega\boldsymbol{w}(t)^\perp-
\tfrac{1}{2}\omega^2\tau\alpha\tau
(1+\alpha\tau)^{-2}\boldsymbol{w}(t)\,.
\end{equation}
More precisely, one can choose an initial condition
$\boldsymbol{w}(0)$, such that
$\boldsymbol{w}(0) = \overline{\boldsymbol{v}}+\mathcal{O}(\omega\tau)$ and
\begin{equation}\label{1.10a}
|\boldsymbol{v}(t)-\boldsymbol{w}(t)|\leq\mathcal{O}\big((\omega\tau)^2\big)
\end{equation}
for all $t/\tau \geq|\log(\omega\tau)^2/\log \gamma_{\alpha\tau}|$. Note that there is a slip
in the initial velocity to
$\overline{\boldsymbol{v}}+\mathcal{O}(\omega\tau)$. In the physical
regime, instead of the more complicated memory equation (\ref{1.2}),
one can work with the friction differential equation (\ref{1.10}),
which in units of (\ref{1.2}) reads
\begin{equation}\label{1.11}
m_\mathrm{eff}\frac{d}{dt}\boldsymbol{v}(t)=\frac{e B_0}
 c\boldsymbol{v}(t)^\perp-\frac{e^2} {6\pi
c^3}\Big(\frac{e B_0} {cm_\mathrm{eff}}\Big)^2 \boldsymbol{v}(t)\,.
\end{equation}

A detailed discussion and derivation of these results is presented
in Sections 2 and 3, where we also explain the point charge limit
$R\to 0$. In our context Landau and Lifshitz \cite{LL} provide a
simple recipe how to proceed from (\ref{1.5a}) to arrive directly at
(\ref{1.11}). This point will be discussed in more detail in Section
4. Based on the novel insights from the Sommerfeld-Page equation we
return to the full model, a charge coupled to its Maxwell
self-field, and summarize of what has been achieved and what still
has to be done, in our opinion.


\section{Stability of the memory equation}\label{sec2}
 \setcounter{equation}{0}

 For the study of the Sommerfeld-Page equation it will be convenient to use complex notation. Then
 $v(t)=v_1(t)+\mathrm{i}v_2(t)\in\mathbb{C}$ and
 $u(t)=u_1(t)+\mathrm{i}u_2(t)\in\mathbb{C}$. With this notation
 (\ref{1.4}) reads
 \begin{equation}\label{2.0}
 \frac{d}{dt}v(t)=-\mathrm{i}\omega v(t)+\alpha\big(v(t-\tau)-v(t)\big)
 \end{equation}
 and, according to (\ref{1.5}),
\begin{equation}\label{2.1}
v(t)=u(t)\quad \textrm{ for }\quad 0\leq t\leq \tau.
\end{equation}
To find out $v(\tau+t)$, we integrate (\ref{2.0}) on both sides.
Then
\begin{equation}\label{2.2}
v(\tau+t)= \mathrm{e}^{-(\alpha+ \mathrm{i}\omega) t} u(\tau) +\alpha\int^t_0
\mathrm{d}s\mathrm{e}^{-(\alpha + \mathrm{i}\omega)(t-s)} u(s)=(K_\omega u)(t)\,,
\end{equation}
$0\leq t\leq \tau$. We regard (\ref{2.2}) as a linear map which out
of $v$ in the interval $[0,\tau]$ makes $v$ in the interval
$[\tau,2\tau]$. We will call $K_\omega$ the Sommerfeld-Page
operator. Note that $K_\omega$ is a non-symmetric operator. More
details are explained in Appendix B. To obtain the solution $v(t)$
for all times one has to simply iterate the map (\ref{2.2}) with the
result
\begin{equation}\label{2.8}
v(n\tau+t)= \big((K_\omega)^n u\big)(t)\,,
\end{equation}
$n$ integer and $0\leq t\leq\tau$. Clearly, $(K_\omega u)(0) = u(\tau)$. If $u$ is arbitrary, then $v(t+\tau)$, $0 \leq t \leq \tau$, is continuous  and so is  $v(t)$ for all $\tau \leq t < \infty$ by iteration. 

The stability of solutions is determined by the eigenvalue problem
for $K_\omega$,
\begin{equation}\label{2.9}
K_\omega u=\lambda u\,.
\end{equation}
Stable dynamics is equivalent to $|\lambda|\leq 1$ for all
eigenvalues. Differentiating (\ref{2.9}) with respect to $t$ yields
\begin{equation}\label{2.7a}
\lambda\frac{d}{dt}u(t)=\alpha u(t) - \lambda(\alpha +\mathrm{i}\omega)
u(t)\,,
\end{equation}
which implies that the eigenvector $u$ is an exponential function.
Setting
\begin{equation}\label{2.10}
u(t)=\mathrm{e}^{z t/\tau} \,,
\end{equation}
one concludes
\begin{equation}\label{2.11}
\lambda=\mathrm{e}^z
\end{equation}
and $|\lambda|\leq 1$ precisely if $\Re z\leq 0$. Inserting in
(\ref{2.7a}) one obtains
\begin{equation}\label{2.12}
z= -\mathrm{i}\omega\tau +\alpha\tau(\mathrm{e}^{-z} -1)\,.
\end{equation}
We will study case by case and, to simplify notation, switch to
dimensionless variables. Since we want to allow $m_0<0$, it is
convenient to set $\tau=1$ for a while.\medskip\\
(i) \textit{The case $\omega=0$.} The eigenvalues of $K_0$ are
determined by
\begin{equation}\label{2.14}
z=\alpha(\mathrm{e}^{-z}-1)\,.
\end{equation}
We set $z=x+\mathrm{i}y$, $x,y\in\mathbb{R}$, and first study
solutions for which $\Im z=0$. They are determined by
\begin{equation}\label{2.10a}
x+\alpha=\alpha\mathrm{e}^{-x}\,.
\end{equation}
$x=0=x_0(\alpha)$ is always a solution. Let us introduce the
two critical points, $\alpha_+$, $\alpha_-$.
$\alpha_-=-1$ and $\alpha_+$ is determined by
\begin{equation}\label{2.17}
\alpha_+
\mathrm{e}^{\alpha_+}=\mathrm{e}^{-1}\,,\quad
\alpha_+\cong 0.278\,.
\end{equation}
(\ref{2.10a}) has one further solution, $x_1(\alpha)$, provided
$\alpha<0$, where $x_1(\alpha)<0$ for $\alpha_-<\alpha<0$ and
$x_1(\alpha)>0$ for $\alpha<\alpha_-$. Thus the motion is
unstable for $m_{\mathrm{eff}}/m_0<0$.

Equating absolute value and phase of (\ref{2.14}), one arrives at
\begin{align}\label{2.15}
(\mathrm{A}) &\qquad (x+\alpha)^2 +y^2=\alpha^2 \mathrm{e}^{-2x}\,,\\
(\mathrm{P}) &\qquad\quad  x+\alpha=-y\cot y\,.
\end{align}
The intersections of the curves (A) and (P) define the additional eigenvalues. (P) has
branches labeled by $m\in\mathbb{Z}$ corresponding to
$(m-1)\pi<y<m\pi$, $m\leq -1$, $-\pi<y<\pi$, $m\pi<y<(m+1)\pi$,
$m>0$. The 0-th branch has its apex at $x=-1-\alpha$, $y=0$ and
converges to $\pm\pi$ as $x\to\infty$. The other branches are
strictly increasing for $m>0$, strictly decreasing for $m<0$, and
cover the full real line, see Fig.1.

\begin{figure}[!htb]
\centering
\includegraphics[width=4.8cm, height=3.5cm]{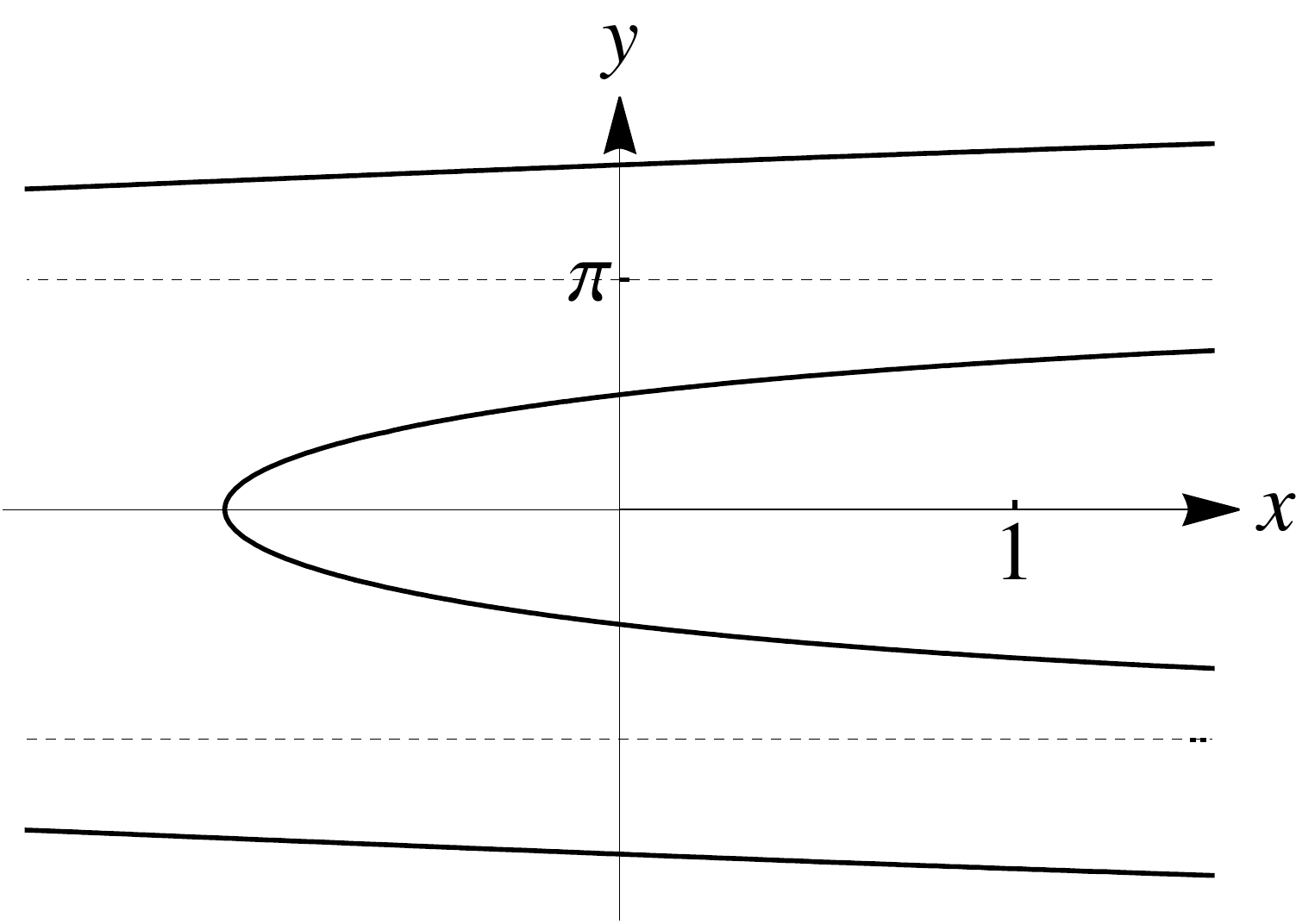}
\caption{Curve (P) for $\alpha = 0$. Eigenvalues result by intersection with one of the curves from Fig. 2, resp. Fig. 3.}
\end{figure}

\begin{figure}[!htb]
\centering
\subfloat{\includegraphics[width=4.0cm, height=3.5cm]{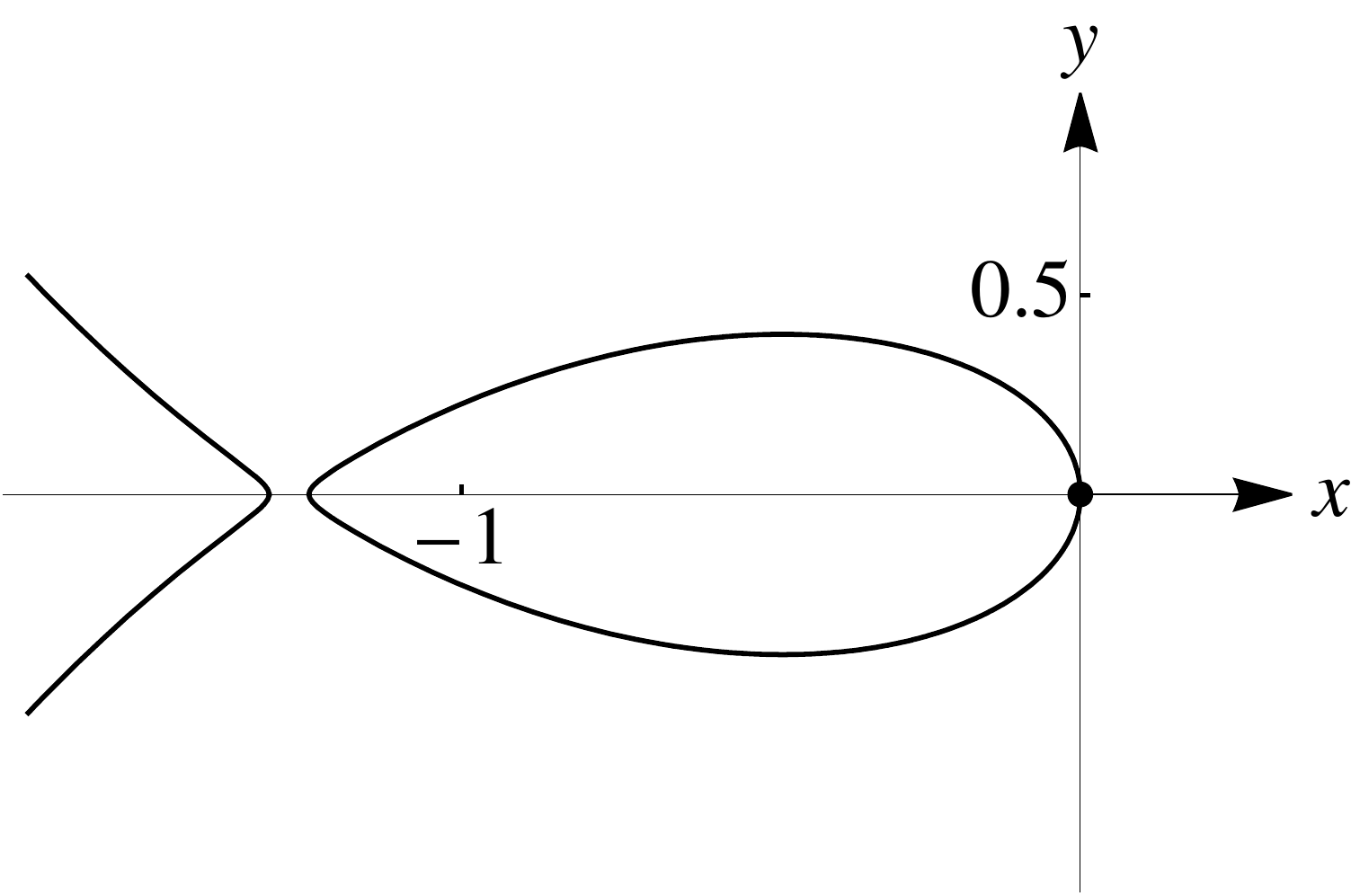}}
\hspace{15mm}     \subfloat{\includegraphics[width=3.8cm,
height=3.5cm]{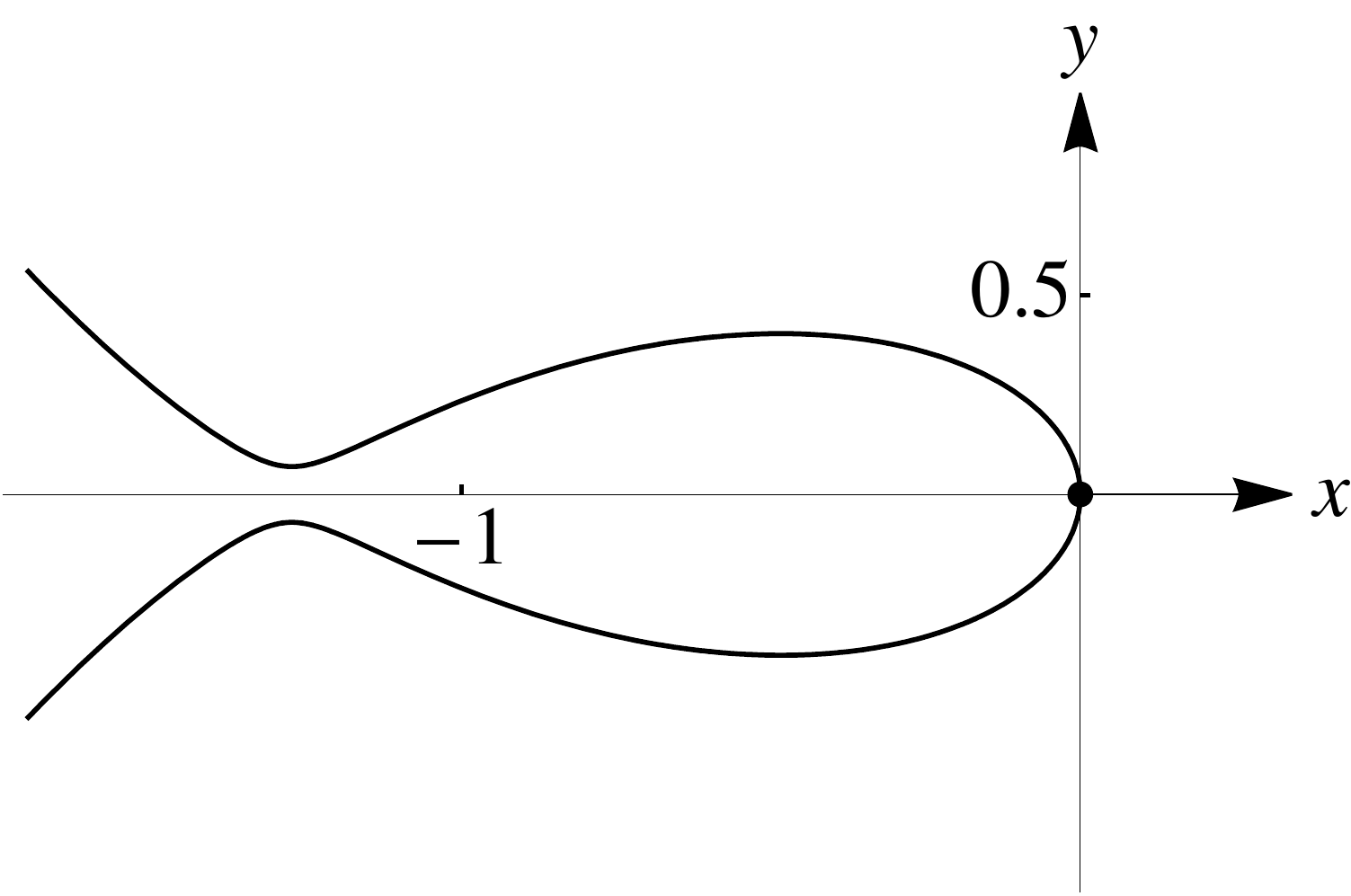}} \hspace{0mm}
\caption{ (a) the curve (A) for
$\alpha =  0.27835 < \alpha_\mathrm{c}^+$ and (b) for $\alpha = 0.279 > \alpha_\mathrm{c}^+$. $\bullet$ is an eigenvalue.}
\end{figure}

\begin{figure}[!htb]
\centering
\subfloat{\includegraphics[width=4.0cm, height=3.5cm]{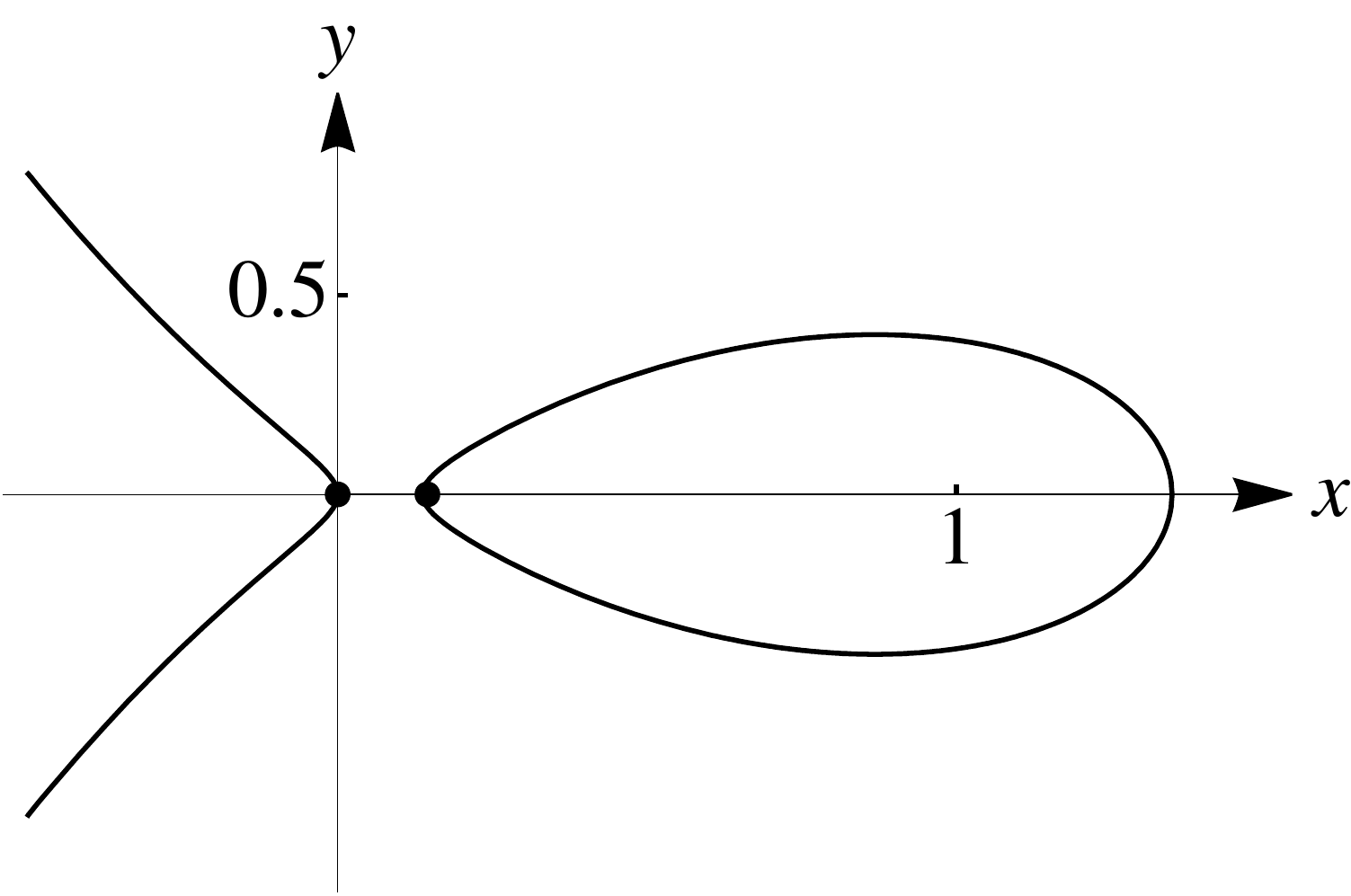}}
\hspace{15mm}     \subfloat{\includegraphics[width=4.8cm,
height=3.5cm]{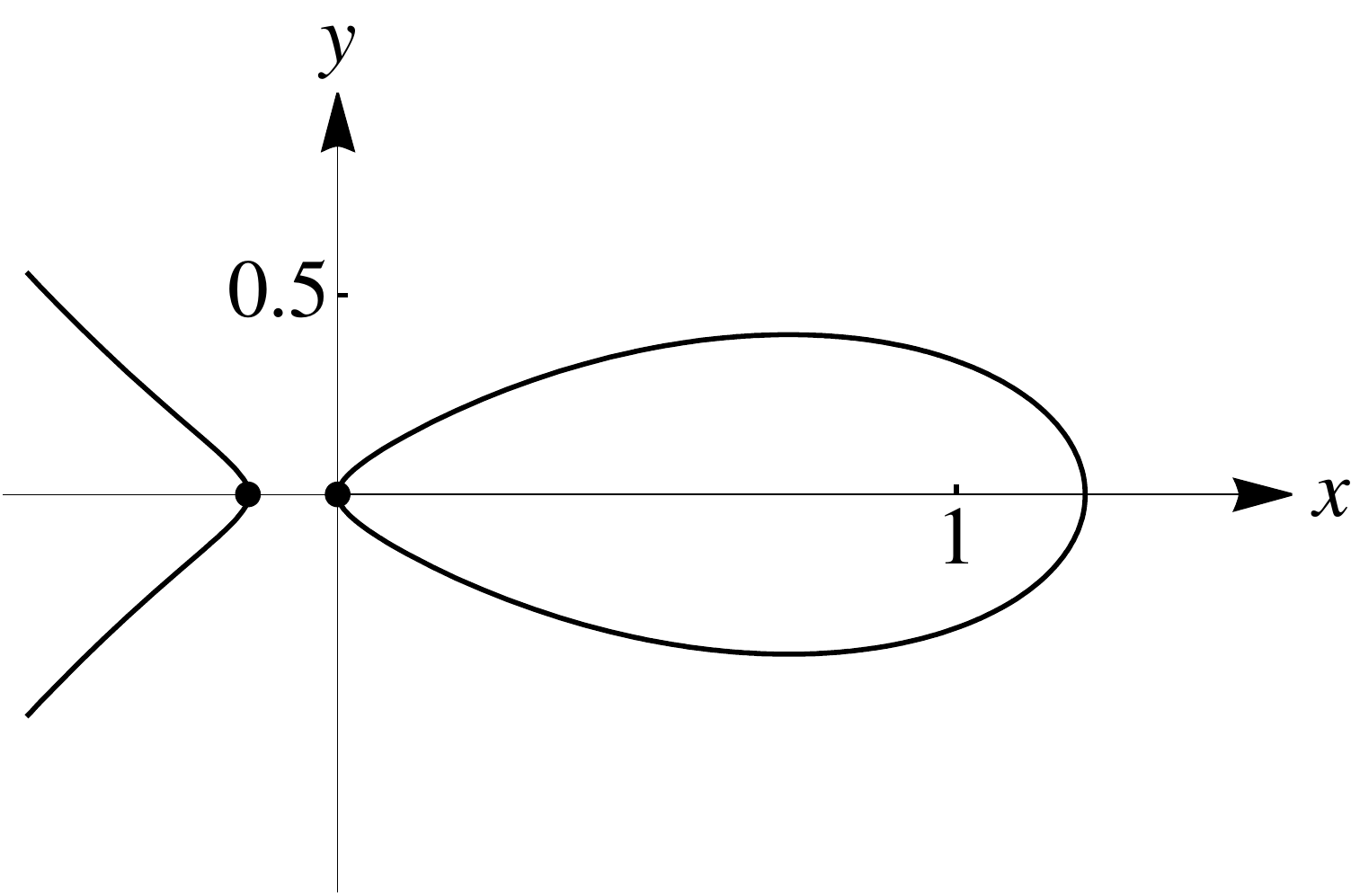}} \hspace{0mm}
\caption{(a) the curve (A) for
$\alpha =  -1.07 < \alpha_\mathrm{c}^-$ and (b) for $\alpha = -0.93> \alpha_\mathrm{c}^-$. $\bullet$ is an eigenvalue.}
\end{figure}


Next we discuss (A),  which has the following
solutions at $y=0$, see Fig. 2 and 3: $x_0(\alpha)=0$ is a solution for all $\alpha$.
$x_1(\alpha)$ and $x_2(\alpha)$ exists only for
$\alpha<\alpha_+$. One has $x_1(\alpha)<x_2(\alpha)<0$
for $0<\alpha<\alpha_+$, $x_1(\alpha)<0<x_2(\alpha)$ for
$\alpha_-<\alpha<0$, and $0<x_1(\alpha)<x_2(\alpha)$ for
$\alpha<\alpha_-$. (A) has always a branch which diverges
to $\pm\infty$ as $x\to-\infty$. Its apex is located at
$(x_1(\alpha),0)$ for $\alpha_-<\alpha < \alpha_+$ and at $(0,0)$
for $\alpha<\alpha_-$, $\alpha_+<\alpha$. In addition (A) has a bubble with
apices at $(x_2(\alpha),0)$ and $(0,0)$ for
$\alpha_- <\alpha<\alpha_+$ and at
$(x_1(\alpha),0)$ and $(x_2(\alpha),0)$ for $\alpha<\alpha_-$.

The 0-th branch of (P) never intersects (A). The branches $m\neq 0$
of (P) intersect the diverging branch of (A), which thus yields only
stable eigenvalues. For $0<\alpha<\alpha_+$ they satisfy
$\Re z<-1-\alpha_+$. The spectrum of eigenvalues of
$K_{\omega}$ at $\omega = 0$ is plotted schematically in Fig. 4. The
decay constant $\gamma_{\alpha\tau}$ is the radius of the smallest
disk which contains all eigenvalues except for the eigenvalue 1. For
$\alpha<\alpha_-$ the intersection points lie to the left
but very close to the imaginary axis. Hence, the stability of motion is
determined by the solutions to (\ref{2.10a}).
\begin{figure}[!htb]
\centering
\includegraphics[width=5cm, height=5cm]{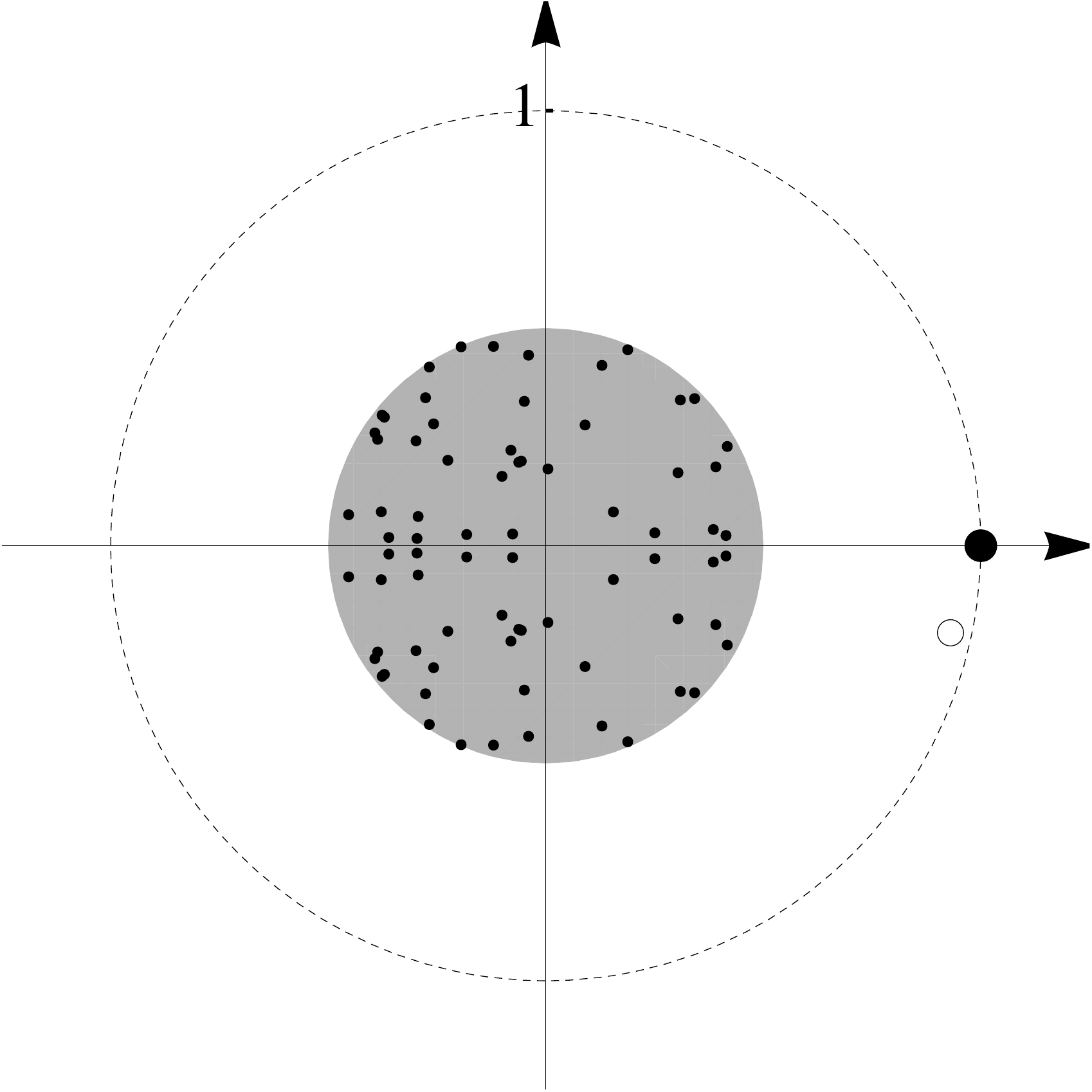}
\caption{Schematic spectrum of $K_\omega$. $\bullet$ are eigenvalues
for $\omega = 0$. The radius of the shaded disk is the decay
constant. $\circ$ indicates the shift of the eigenvalues by taking
$\omega \neq 0$.}
\end{figure}
\medskip\\
(ii) \textit{The case $\omega\neq 0$, $|\omega|\ll 1$.} The
eigenvalues are determined by
\begin{equation}\label{2.18}
z=-\mathrm{i} \omega +\alpha(\mathrm{e}^{-z}-1)\,,
\end{equation}
which induces a small shift. Of interest is the shift of the
eigenvalue $\lambda=1$, i.e. $z=0$. Expanding at $z=0$, one arrives
at
\begin{equation}\label{2.19}
z=-\mathrm{i}\omega+\alpha (-z+\tfrac{1}{2}z^2)\,,
\end{equation}
which for small $\omega$ has the solution
\begin{equation}\label{2.20}
z= -\frac{\alpha}{2(1+\alpha)^3}\omega^2-\mathrm{i}
\frac{1}{(1+\alpha)}\omega\,.
\end{equation}
For $\alpha>0$ the motion is stable, while
for $-1<\alpha<0$ it is unstable. At $\alpha = - 1$ there is a double degenerate eigenvalue $0$,
which splits into the solutions of $z^2 = -2\mathrm{i}\omega$. Hence the unstable motion persists for $\alpha \leq - 1$.
For $|\omega|\ll 1$ and $\alpha >0$, there is a single maximal eigenvalue, $\lambda_\mathrm{max}$, with the property
\begin{equation}\label{2.23}
|\lambda_\mathrm{max}|=1-\frac{1}{2}\frac{\alpha}{1+\alpha}
\big(\frac{\omega}{1+\alpha}\big)^2 + \mathcal{O}(\omega^3)\,,
\end{equation}
which implies a slow inward motion.

\section{Long time behavior and comparison dynamics}\label{sec3}
 \setcounter{equation}{0}

 These two topics are not specific for a charge coupled to the Maxwell
 field. The long time behavior
 is a first central task in the analysis of a dynamical system. Comparison dynamics is the
 issue of approximating the full dynamics through simplified, more tractable equations
 and typically requires to have a small parameter at one's disposal. The
 generic form of such a comparison is to first match the initial
 data then to provide error bars which specify how well and over
 what time scale a solution of the true dynamics is shadowed by the
 comparison dynamics. In the following we denote by $v(t)$ the true
 dynamics, as before, and the approximation by $w(t)$, both taking values in $\mathbb{C}$.

 \subsection{Long time asymptotics}\label{sec3.0}

 To determine the long time behavior we need the maximal eigenvector
 and the spectral gap. As explained in more detail in the appendix, the natural space for
the Sommerfeld-Page operator $K_\omega$ is the Hilbert space
$\mathcal{H}=\mathbb{C}\oplus L^2([0,\tau],dt)$. A vector
$(u_0,u)\in\mathcal{H}$ consists of the constant $u_0=u(\tau)$ and
the function $u(t)$, $0\leq t\leq\tau$. The spectrum of $K_\omega$
acting on $\mathcal{H}$ is denoted by $\sigma(K_\omega)$. As shown
in Appendix B, $K_\omega$ has the family of right eigenvectors
$|g_\lambda\rangle$ and of left eigenvectors $\langle f_\lambda|$,
$\lambda\in\sigma(K_\omega)$, such that
\begin{equation}\label{3.1}
(K_\omega)^n=\sum_{\lambda\in\sigma(K_\omega)}
\lambda^n(1+\lambda^{-1} \alpha\tau)^{-1} |g_\lambda\rangle\langle
f_\lambda|\,.
\end{equation}
Explicit formulae for $|g_\lambda\rangle,\, |f_\lambda\rangle$ are
given in (\ref{A.16}) together with (\ref{A.11}), (\ref{A.12}) at $q=\lambda$.

We assume now $m_0>0$ and $\omega=0$. Then 1 is the unique maximal
eigenvalue with left eigenvector $|g_1\rangle=(1,1)$ and right
eigenvector $|f_1\rangle=(1,\alpha)$. Hence, with
$P_1´=|1+\alpha)^{-1}|g_1\rangle\langle f_1|$,
\begin{equation}\label{3.1a}
(P_1 u)(t)=(1+\alpha \tau)^{-1} \big(u(\tau)+\alpha\int^{\tau}_0 \mathrm{d}s
u(s)\big)=\overline{v}
\end{equation}
for all $t$, $0\leq t\leq\tau$, which equals Eq. (\ref{1.7}) in complex rather than
2-vector notation. The estimates for the decay constant are derived
from the graphical representation of Section 2.

Turning to the case $\omega\neq 0$, the maximal eigenvalue
$\lambda_\mathrm{max}$ has the property that
$|\lambda_\mathrm{max}|<1$. Therefore $v(t)$ decays to 0 as
$t\to\infty$ with decay constant $|\lambda_\mathrm{max}|$. This
observation yields (\ref{1.9}).


\subsection{Approximation of small $B_0$-field}\label{sec3.1}

We consider the case $|\omega\tau|\ll 1$, which physically corresponds
to a ``weak'' magnetic field and assume $m_0>0$, i.e. $\alpha>0$.
Denoting the integer part by $[\cdot]$, it holds
\begin{equation}\label{3.2}
    v(t)=\big((K_\omega)^{[t/\tau]} u\big)
    \big(t-[t/\tau]\tau\big)\,.
\end{equation}
The maximal eigenvalue of $K_\omega$ is given by the solution to (\ref{2.18}) close to $z = 0$. The
decay constant is $ \gamma_{\alpha\tau}\leq 0.07$. Therefore, with $\lambda= \lambda_{\mathrm{max}}$,
\begin{equation}\label{3.2a}
v(t)=  \lambda^{[t/\tau]} (1 + \lambda^{-1} \alpha \tau)^{-1} |g_\lambda\rangle\langle f_\lambda|u\rangle
   (t-[t/\tau]\tau) + \mathcal{O}((\gamma_{\alpha\tau})^{[t/\tau]})\,.
\end{equation}
For $|\omega \tau| \ll  1$, one obtains
\begin{equation}\label{3.5}
\lambda^{[t/\tau]} \to
\exp\big[\big(-\mathrm{i}(1+\alpha\tau)^{-1} \omega-
\alpha\tau(2(1+\alpha\tau)^3)^{-1} \omega^2 \tau\big)t\big] + \mathcal{O}((\omega \tau)^2)\,.
\end{equation}
The initial value is  $\overline{\boldsymbol{v}}+\mathcal{O}(\omega\tau)$. Finally, if $t/\tau \geq|\log(\omega\tau)^2/\log \gamma_{\alpha\tau}|$, then
$(\gamma_{\alpha\tau})^{[t/\tau]} \leq (\omega \tau)^2$. Thereby we have established (\ref{1.10}) and (\ref{1.10a}).


\subsection{The point charge limit}\label{sec3.2}

Following Dirac \cite{Di} we consider the limit
\begin{equation}\label{3.8}
\tau\to 0 \quad \textrm{for fixed}\quad
m_\mathrm{eff}=m_0+\frac{e^2}{3\pi c^3\tau}>0\,.
\end{equation}
Then, for small $\tau$,
\begin{equation}\label{3.9}
\alpha\tau=-1-a\tau\,,\quad a=\frac{3\pi c^3}{e^2}m_\mathrm{eff}\,,
\end{equation}
and
\begin{equation}\label{3.10}
\omega\tau=-b\tau^2\,,\quad b=\frac{3\pi c^3}{e^2}\frac{e B_0}{c}\,.
\end{equation}
Since $\alpha\tau<-1$, but close to $-1$, we now have to take into
account the two maximal eigenvalues, $\lambda_+$, $\lambda_-$, close to 1. The
eigenvalue equation reads
\begin{equation}\label{3.11}
z +\mathrm{i}\omega\tau = \alpha\tau (e^{-z}-1)\,.
\end{equation}
Setting $z=\kappa\tau$ one finds
\begin{equation}\label{3.12}
\lambda_\pm= 1+\tau\kappa_\pm\,,\quad \kappa_\pm =-a\big(1\pm
\sqrt{1+\mathrm{i}2b/a}\,\big)\,.
\end{equation}
In the limit $\tau\to 0$ it holds
\begin{equation}\label{3.13}
(1+\tau\kappa_\pm)^{[t/\tau]}\to \mathrm{e}^{\kappa_\pm t}\,.
\end{equation}
In differential form, (\ref{3.13}) agrees with Lorentz-Dirac
equation (\ref{1.11}).

Somewhat more tricky is the computation of the initial conditions.
In the sum (\ref{3.1}) only the eigenvalues $\lambda_\pm$
contribute. All other eigenvalues satisfy $|\lambda|\leq 0.25$,
independent of $\tau$. We set $\lambda=1+\tau\kappa$ and omit the
$\pm$ index. For given $u(t)$, $0\leq t\leq\tau$, one has to study
\begin{equation}\label{3.14}
|g_\lambda\rangle\langle f_\lambda|u\rangle
(1+\alpha\tau\lambda^{-1})^{-1}
\end{equation}
in the limit $\tau\to 0$.

Since, according to (\ref{A.12}), (\ref{A.16}),
\begin{equation}\label{3.15}
|g_\lambda\rangle =\big(1,z^{-1} \exp[\alpha t(z^{-1}
-1)-\mathrm{i}\omega t]\big)\,,
\end{equation}
with $\lambda=\mathrm{e}^z$, one finds
\begin{equation}\label{3.16}
\lim_{\tau\to 0} |g_\lambda\rangle =(1,1)\,.
\end{equation}

The left eigenvector $|f_\lambda\rangle$ is given by
\begin{equation}\label{3.17}
|f_\lambda\rangle=(1,f_\lambda (t))
\end{equation}
with $\lambda=\mathrm{e}^z$ and
\begin{eqnarray}\label{3.18}
&&\hspace{-20pt}
f_\lambda(t)=-\big(1-2\mathrm{i}\omega(z/\alpha)\big)^{-1}
2\mathrm{i}\omega e^{-(\alpha-\mathrm{i}\omega)(\tau-t)}\nonumber\\
&&\hspace{20pt}
+\big(1-2\mathrm{i}\omega(z/\alpha)\big)^{-1}(\alpha/z) \exp[\alpha
t(z^{-1} -1) -\mathrm{i}\omega(\tau-t)]\,.
\end{eqnarray}
Hence the scalar product becomes
\begin{equation}\label{3.19}
\langle f_\lambda|u\rangle = u(\tau) + \int^\tau_0 \mathrm{d}s u(s)
f_\lambda(s)^\ast\,.
\end{equation}
The contribution from the first summand of (\ref{3.18}) is
$\mathcal{O}(\tau^3)$. For the second term we expand as $u(t)=u(0)
+u'(0)t$ and obtain the scalar product of (\ref{3.19}) as
$\frac{1}{2}\tau(u'(0)-\kappa u(0))$. Hence
\begin{equation}\label{3.20}
\lim_{\tau\to 0} |g_\lambda\rangle  \langle f_\lambda |u\rangle
(1+\alpha\tau\lambda^{-1})^{-1}=\big(2(\kappa-a)\big)^{-1}
\big(u'(0)-\kappa u(0)\big)\,.
\end{equation}
We conclude that in the point charge limit $v(t) \to w(t)$ as $\tau \to 0$ constrained by 
(\ref{3.8}) and the limit motion is given by
\begin{equation}\label{3.21}
w(t) = \sum_{\sigma=\pm}
\big(2(\kappa_\sigma-a)\big)^{-1}\big(u'(0)-\kappa_\sigma
u(0)\big)\mathrm{e}^{\kappa_\sigma t}\,.
\end{equation}

$\kappa_+$ is the stable and $\kappa_-$ the unstable mode. Unless
there is a specific relation between $u(0)$ and $u'(0)$ the motion
(\ref{3.21}) is unstable. Dirac postulated his asymptotic condition,
which in our context would single out the stable mode. From the
point of view of the memory equation, as soon as $m_0<0$ with
$m_\mathrm{eff}>0$, the motion is unstable. Thus it is not
surprising that also the $\tau\to 0$ limiting motion is unstable. Of
course, one could adjust the initial conditions to be orthogonal to
$|f_{\lambda_-}\rangle$, thus maintaining a stable solution
throughout the limiting procedure. Clearly, such a constraint would
have to be justified by some additional reasoning beyond the
Sommerfeld-Page equation.
\section{The Landau-Lifshitz approximation}\label{sec4}
 \setcounter{equation}{0}

 In the Landau-Lifshitz approximation the external $\boldsymbol{B}_0$-field is
 considered to be small. To introduce the small parameter
 explicitly we substitute in (\ref{1.4}) $\omega$ by
 $\varepsilon\omega$ with $0 < \varepsilon\ll 1$ a dimensionless
 parameter. To make up for the small field one has to consider times
 of order $\varepsilon^{-1}$. The rescaled time, again denoted by
 $t$, is in units of single revolutions. With both changes
 (\ref{1.4}) becomes
 \begin{equation}\label{4.1}
 \varepsilon\frac{d}{dt}\boldsymbol{v}(t) = \varepsilon \omega\boldsymbol{v}(t)^\perp + \alpha
 \big(\boldsymbol{v}(t-\varepsilon\tau)-\boldsymbol{v}(t)\big)\,.
 \end{equation}
 Taylor expanding in $\varepsilon$ yields
 \begin{equation}\label{4.2}
 (1+\alpha\tau)
 \frac{d}{dt}\boldsymbol{v}(t)=\omega\boldsymbol{v}(t)^\perp
 +\tfrac{1}{2}\alpha\tau^2
 \varepsilon\frac{d^2}{dt^2}\boldsymbol{v}(t)\,.
 \end{equation}
The small parameter $\varepsilon$ appears now in front of the
highest derivative, which is the hallmark of singular perturbation
theory \cite{J}. The center manifold to order $\varepsilon$ is singled out by
re-substituting the leading order in (\ref{4.2}) with the result
\begin{equation}\label{4.3}
 (1+\alpha\tau)
 \frac{d}{dt}\boldsymbol{v}(t)=\omega\boldsymbol{v}(t)^\perp
 -\tfrac{1}{2}\varepsilon\alpha (\frac{\omega\tau}{1+\alpha\tau})^2
 \boldsymbol{v}(t)\,.
 \end{equation}

 It seems that by a clever short-cut we have directly arrived at
 (\ref{1.10}). If one compares with (\ref{1.5a}), then  (\ref{4.3}) corresponds to
 the weak $B_0$-field approximation of the Lorentz-Dirac equation
 restricted to the stable plane $\mathcal{C}_\omega$ of physical solutions.

 If one tries to control the error in the step from (\ref{4.1}) to
 (\ref{4.2}), one runs into the same difficulty as with the
 Lorentz-Dirac equation. Unstable solutions increase as
 $\mathrm{e}^{t/\varepsilon}$ and it remains mysterious by which fine
 tuning of the initial conditions such runaways are avoided. But
 since we know from Section 3 that (\ref{4.3}) is a controlled
 approximation to the true solution, we conclude that Taylor
 expanding (\ref{4.1}) generates spurious solutions. They are an
 artifact of the approximation scheme used and not an intrinsic property of the
 Sommerfeld-Page equation.

\section{Discussions and conclusions}\label{sec5}
\setcounter{equation}{0}

Physically one would expect that the results from the case study are
of a much greater generality. Of course the goal would be to treat a
fully relativistic matter field coupled to the Maxwell field. Not so many
rigorous results are available then and we propose to first have a
look at the semi-relativistic Abraham model.

The dynamical variables are position $\boldsymbol{q}$, the velocity $\boldsymbol{v}$, and the
self-fields $\boldsymbol{E}(\boldsymbol{x},t)$,
$\boldsymbol{B}(\boldsymbol{x},t)$. In addition there is a uniform
external magnetic field $\boldsymbol{B}_0$. Then the coupled
equations of motion read
\begin{eqnarray}\label{5.1}
&&c^{-1} \frac{\partial}{\partial t}\boldsymbol{B}=-\nabla \times
\boldsymbol{E}\,, \quad
c^{-1} \frac{\partial}{\partial t} \boldsymbol{E}= \nabla \times
\boldsymbol{B} -\frac{e}{c}\varphi(\cdot -q) \boldsymbol{v}\,,\\[1 ex]
&&
\nabla \cdot \boldsymbol{E} = e\varphi (\cdot -q)\,,\quad
\nabla\cdot \boldsymbol{B}=0\,,\\[1 ex]\label{5.4}
&&\hspace{-30pt}
\frac{d}{dt}m_\mathrm{b}
\gamma\boldsymbol{v}(t)=\frac{e}{c}
\boldsymbol{v}(t)   \times\boldsymbol{B}_0+\frac{e}{c}\int \mathrm{d}x
\varphi(\boldsymbol{x}-\boldsymbol{q}(t)) \big(
\boldsymbol{E}(\boldsymbol{x},t)+\boldsymbol{v}(t)\times
\boldsymbol{B}(\boldsymbol{x},t)\big)\,.
\end{eqnarray}
Here $\varphi$ is the rigid charge distribution, $\varphi\geq 0$,
$\varphi$ radial and smooth, $\varphi(\boldsymbol{x})=0$ for
$|\boldsymbol{x}|\geq R$, $\int \mathrm{d}x \varphi(\boldsymbol{x})=1$.
$m_\mathrm{b}$ is the mechanical bare mass and
$\gamma(\boldsymbol{v})=(1-(\boldsymbol{v}/c)^2)^{-1/2}$. The initial position
and velocity for the particle is
$\boldsymbol{q}_0,\boldsymbol{v}_0,|\boldsymbol{v}_0/c|<1$. For the
self-field there is some freedom, as discussed already. To be
specific we set
\begin{equation}\label{5.5}
\boldsymbol{E}(\boldsymbol{x},0)=\boldsymbol{E}_{\boldsymbol{v}_0}
(\boldsymbol{x}-\boldsymbol{q}_0)\,,\quad
\boldsymbol{B}(\boldsymbol{x},0)=\boldsymbol{B}_{\boldsymbol{v}_0}
(\boldsymbol{x}-\boldsymbol{q}_0)\,,
\end{equation}
where $\boldsymbol{E}_{\boldsymbol{v}_0}$,
$\boldsymbol{B}_{\boldsymbol{v}_0}$ are the minimizers of the total, i.e. field
plus particle, energy at given $\boldsymbol{q}_0,\boldsymbol{v}_0$.
For an explicit formula we refer to  \cite{Sp}, Eqs. (4.5) to
(4.7).

As for the Sommerfeld-Page equation, the motion in the plane orthogonal 
to $\boldsymbol{B}_0$ can be decoupled. For notational simplicity, let us set $\boldsymbol{B}_0 = (0,0,B_0)$. We assume 
\begin{equation}\label{5.5a}
q_3(0) = 0\,,\quad v_3(0) = 0
\end{equation}
and require for the initial fields that
\begin{equation}\label{5.5b}
E_3(x_1,x_2,x_3) = - E_3(x_1,x_2,-x_3)\,,\quad B_j(x_1,x_2,x_3) = 
-B_j(x_1,x_2,- x_3)\,,\quad j = 1,2\,.
\end{equation}
Then from (\ref{5.1}) -- (\ref{5.4}) it follows that these properties remain valid for all $t$.
Thus for initial fields with the particular symmetry (\ref{5.5b}) the motion of the charge stays in the 1-2 plane.

We now follow the scheme of Section 4. In the fully relativistic
setting this yields the effective equations of motion
\begin{equation}\label{5.6}
m_0\gamma\frac{d}{dt}\boldsymbol{v}=\frac{e}{c}(\boldsymbol{v}\times
\boldsymbol{B}_0)+\frac{e^2}{6\pi c^3 m_0} (\frac{e}{m_0 c})^2
\big[(\boldsymbol{v}\cdot \boldsymbol{B}_0)
\boldsymbol{B}_0-\boldsymbol{B}^2_0 \boldsymbol{v}\big]
\end{equation}
valid for ``small" $\boldsymbol{B}_0$. Here $m_0$ is the rest mass of
the particle. If one would aim at a precision of order $|\boldsymbol{B}_0|$,
then the initial conditions for (\ref{5.6}) are simply
$\boldsymbol{q}_0$, $\boldsymbol{v}_0$.

The semi-relativistic character of Abraham model is merely reflected in a velocity dependent effective mass
which
differs from the one of a relativistic particle. The relevant computation can be found in
\cite{Sp}, Chapter 4.1. With this difference one can still work
out the re-substitution as indicated in Section 4. The expressions become
unwieldy and there is no particular reason to display them here. The structure
of the effective equations of motion is similar to (\ref{5.6}). 

For our particular initial condition, if $\boldsymbol{B}_0=0$, then the
particle would simply continue to travel with velocity
$\boldsymbol{v}_0$. Thus one might guess that the effective
equations of motion (\ref{5.6}) should be solved with initial
conditions $\boldsymbol{v}(0)=\boldsymbol{v}_0$. As discussed at length in our case
study, to have an approximation which includes friction and thus allows for an error
$\mathcal{O}(\boldsymbol{B}^2_0)$ only, one expects that the initial
conditions must be corrected slightly by a term of order $|\boldsymbol{B}_0|$.

The details of such a  program are carried out in  \cite{KS}. As emphasized already, upon  Taylor expanding 
one arrives at a comparison dynamics which 
has runaway solutions making  difficult the precise control of errors.
The best we could accomplish is an estimate of the difference in energies of the true and
comparison dynamics. A pointwise comparison of the trajectories is
missing.

More tractable is the issue of the long time asymptotics. Under
certain conditions it is proved that, in case $\boldsymbol{B}_0 = 0$, 
$\lim_{t\to \infty}\boldsymbol{v}(t)=\boldsymbol{\overline{v}}$ \cite{IKS1,IKM,IKS2}.
Also, if the charge is subject to an external electrostatic potential, then 
in for long times the particle
comes to rest at a critical point of the potential  \cite{KoSp}. On the other side, for a uniform external magnetic field 
$\boldsymbol{B}_0 \neq 0$ the energy balance used in \cite{KoSp} is not strong enough to control the long time behavior and
it remains as an open problem to establish that the charge 
comes to rest in the limit $t\to\infty$. 
\bigskip\\
\textbf{Acknowledgements}. 
I thank D. Griffiths for most helpful comments on a preliminary version of this paper.
I am grateful to D. Giulini and V. Perlick as organizers of the 475. WE-Heraeus-Seminar  
``Problems and Developments of Classical Electrodynamics''. The exchanges at the workshop 
strongly motivated me to finish up my notes. In this context discussions with A. Harte, J. Kirjowski, R. Wald, and A. Yaghjian
are acknowledged. 
I thank Martin F\"{u}rst for most useful interactions
 and for help with the Figures.

\begin{appendix}
\section{Appendix: Derivation of the Sommerfeld-Page equation}\label{app.B}

In standard derivations of the Sommerfeld-Page equation, one
expands the solution of the full
Abraham model in $R$ as a small parameter retaining only the terms linear in
$\boldsymbol{v}$. In \cite{Sp}, Chapter 7, we provide a
somewhat different perspective. No originality is claimed.
Starting from the full Abraham model one can write down an exact
memory equation for the motion of the charge. There are two terms:
one term results from the initial conditions of the Maxwell field
and a memory term, which contains only the self-generated field. The
charge distribution is assumed to be radially  symmetric and
supported in a ball of radius $R$, but otherwise arbitrary, and the Maxwell field is 
assumed to be co-moving.
Then memory refers to times up to $2R/c$ into the past and after a
time of order $2R/c$ the initial Maxwell field no longer influences
the motion of the charge. No expansion is employed. If one
now specializes to a charge distribution uniformly over a sphere of
radius $R$ and retains in the memory term only contributions linear
in $\boldsymbol{v}$, then one arrives at the Sommerfeld-Page
equation (\ref{1.2}).

\section{Appendix: The Sommerfeld-Page operator}\label{app.A}
\setcounter{equation}{0}

We provide a more detailed analysis of the Sommerfeld-Page operator
$K_\omega$ which reads
\begin{equation}\label{A.1}
(K_\omega u)(t)=\mathrm{e}^{-(\alpha + \mathrm{i}\omega)t}u(\tau)+\alpha\int^t_0
\mathrm{d}s\mathrm{e}^{-(\alpha+\mathrm{i}\omega)(t-s)} u(s)\,,\quad
0\leq t\leq \tau\,.
\end{equation}

The eigensolutions to (\ref{A.1}),
\begin{equation}\label{A.2}
    K_\omega u_\lambda=\lambda u_\lambda\,,
\end{equation}
are of the form
\begin{equation}\label{A.3}
u_\lambda(t)=\mathrm{e}^{z t/\tau}\,,\quad \lambda=\mathrm{e}^z\,.
\end{equation}
Here $\lambda\in\sigma(K)$, the spectrum of $K$, and $\lambda$ is
implicitly determined through the solutions of
\begin{equation}\label{A.3a}
z+\mathrm{i}\omega\tau=\alpha\tau(\mathrm{e}^{-z}-1)\,,
\end{equation}
which coincides with (\ref{2.12}). Assuming the initial condition $u$ as the
linear superposition
\begin{equation}\label{A.4}
u(t)=\sum_{\lambda\in\sigma(K_\omega)} c_\lambda u_\lambda(t)\,,
\end{equation}
one finds
\begin{equation}\label{A.5}
(K_\omega)^n u(t)=\sum_{\lambda\in\sigma(K_\omega)}c_\lambda
\lambda^n u_\lambda(t)\,.
\end{equation}
For the particular case $\omega=0$ and $1+\alpha\tau>0$, $\lambda=1$
is the unique maximal eigenvalue. We then split as
\begin{equation}\label{A.5a}
(K_0)^n u(t)=c_1 +\sum_{\lambda\in\tilde{\sigma}(K_\omega)}c_\lambda
\lambda^n u_\lambda(t) \,,
\end{equation}
$\tilde{\sigma}(K)=\sigma(K)\setminus\{1\}$. Therefore
\begin{equation}\label{A.6}
|(K_0)^n u(t)-c_1|\leq
\big(\sum_{\lambda\in\tilde{\sigma}(K)}|c_\lambda|\big) \gamma^n
\end{equation}
with the decay constant
$\gamma=\mathrm{max}\big\{|\lambda|,\lambda\in\tilde{\sigma}(K)\big\}$.
While this provides the desired exponential decay, the constant in
front of $\gamma^n$ might be large. In particular, one would like to have a bound
which is expressed directly in terms of the initial $u$ and not in
the expansion coefficients $c_\lambda$.

If in (\ref{A.1}) $u$ is assumed to be continuous, then the dual
space is the space of finite complex measures over $[0,\tau]$. After
one application of the dual operator $K^{\star}_\omega$ the measure has a
continuous part $u(t) \mathrm{d}t$, $0\leq t\leq\tau$, and a
$\delta$-function at $\tau$. Thus the natural space for $K_\omega$
is $\mathbb{C}\oplus L^2([0,\tau],\mathrm{d}t)=\mathcal{H}$. For
$(c,u)\in\mathcal{H}$, $c$ corresponds to the complex weight of the
$\delta$-function at $\tau$ and $u$ to $u(t)$, $0\leq t\leq\tau$.
With this representation the operator can be written as
\begin{equation}\label{A.7}
K_\omega= \left(
  \begin{array}{cc}
    \beta & \langle f| \\
    |g\rangle & A \\
  \end{array}
\right)\,,
\end{equation}
where $\beta=\mathrm{e}^{-(\alpha+\mathrm{i}\omega)\tau}$,
$g(t)=\mathrm{e}^{-(\alpha+\mathrm{i}\omega)t}$,
$f(t)=\mathrm{e}^{-(\alpha-\mathrm{i}\omega)(\tau-t)}$, and the
kernel of $A$, $A(t,s)=\chi(s\leq t)\alpha
\mathrm{e}^{-(\alpha+\mathrm{i}\omega)(t-s)}$.

To estimate $(K_\omega)^n$ for large $n$ we use the resolvent of
$K_\omega$. As verified directly by integration, one finds
\begin{equation}\label{A.8}
(q-A)^{-1}(t,s)= q^{-1} \delta(t-s)+q^{-2} \chi(s\leq
t)\alpha\mathrm{e}^{-(\alpha+\mathrm{i}\omega)(t-s)}
\mathrm{e}^{q^{-1}\alpha(t-s)}
\end{equation}
for all $q \in \mathbb{C} \setminus \{0\}$. $K_\omega$ is a one-dimensional perturbation of $A$. Hence
\begin{eqnarray}\label{A.9}
&&\hspace{-50pt}(q-K_\omega)^{-1}= \left(
  \begin{array}{cc}
    (q-\beta)^{-1} & 0 \\
    0 & (q-A)^{-1} \\
  \end{array}  \right)\nonumber\\
&&\hspace{-10pt} +D(q)^{-1}
  \left(
    \begin{array}{cc}
(q-\beta)^{-1} \langle f|(q-A)^{-1}g\rangle  & \langle f| (q-A)^{-1} \\
(q-A)^{-1}|g\rangle & (q-A)^{-1} |g\rangle  \langle f| (q-A)^{-1}\\
    \end{array}
\right)\,,
\end{eqnarray}
where
\begin{equation}\label{A.10}
D(q)= q-\beta -\langle f| (q-A)^{-1} g\rangle\,.
\end{equation}
The scalar product is obtained to
\begin{equation}\label{A.10a}
\langle f|(q-A)^{-1}g\rangle=
\mathrm{e}^{-(\alpha+\mathrm{i}\omega)\tau}
\mathrm{e}^{q^{-1}\alpha\tau}-\mathrm{e}^{-(\alpha+\mathrm{i}\omega)\tau}
\end{equation}
and hence
\begin{equation}\label{A.10b}
D(q)= q-\mathrm{e}^{-(\alpha+\mathrm{i}\omega)\tau}
\mathrm{e}^{q^{-1}\alpha\tau}\,.
\end{equation}
Setting $q=\mathrm{e}^z$ one observes that $D(q)=0$ is equivalent to
(\ref{A.3a}) and (\ref{A.9}) holds whenever $ q \in \mathbb{C} \setminus \sigma(K_\omega)$.

From (\ref{A.8}), using $g(t)$, $f(t)$ as below (\ref{A.7}),

\begin{eqnarray}\label{A.11}
&&\hspace{-55pt}(q-A)^{-1} f(t)=-2\mathrm{i}\omega
(1-2\mathrm{i}\omega q \alpha^{-1})^{-1}
\mathrm{e}^{-(\alpha-\mathrm{i}\omega)(\tau-t)}  \nonumber\\
&&\hspace{32pt} + \alpha q^{-1}
 (1-2\mathrm{i}\omega q \alpha^{-1})^{-1}
\mathrm{e}^{-(\alpha-\mathrm{i}\omega)(\tau-t)}
\mathrm{e}^{q^{-1}\alpha(\tau-t)}=f_q(t)\,,\\[1 ex]
\label{A.12}
&&\hspace{-55pt}(q-A)^{-1} g(t)= q^{-1} \mathrm{e}^{-(\alpha+\mathrm{i}\omega)
t}e^{q^{-1}\alpha t}=g_q(t)\,.
\end{eqnarray}

The poles of the resolvent define the spectrum $\sigma(K_\omega)$.
There is a pole at $q=0$, which does not contribute to the spectral
representation of $K_\omega$. There is an apparent pole at
$q=\beta$, which can be lifted since the numerator vanishes at
$q=\beta$. Finally there are the poles resulting from the zeros of
$D$,
\begin{equation}\label{A.13}
\sigma(K_\omega)=\{\lambda|D(\lambda)=0\}\,,
\end{equation}
which has been shown already to be in agreement with (\ref{A.3a}).
The residue is computed from
\begin{equation}\label{A.14}
D'(\lambda)=1+\tau \lambda^{-1}\,.
\end{equation}
Altogether, with $\mathcal{C}_\lambda$ a small contour around $\lambda$, one
obtains 
\begin{equation}\label{A.15}
P_\lambda= \oint_{\mathcal{C}_\lambda}
\mathrm{d}q(q-K_\omega)^{-1}=(1+\alpha\tau\lambda^{-1})^{-1}
|g_\lambda\rangle\langle f_\lambda|
\end{equation}
for $\lambda\in \sigma(K_\omega)$ with
\begin{equation}\label{A.16}
|g_\lambda\rangle=(1,g_\lambda(t)),\quad
|f_\lambda\rangle=(1,f_\lambda(t))\,.
\end{equation}
With these preparations we obtain
\begin{equation}\label{A.17}
K_\omega=\sum_{\lambda\in\sigma(K_\omega)} \lambda
(1+\alpha\tau\lambda^{-1})^{-1}|g_\lambda\rangle\langle
f_\lambda|\,.
\end{equation}
There are no Jordan block's since the projectors in (\ref{A.17}) are
one-dimensional. As a consequence
\begin{equation}\label{A.18}
(K_\omega)^n=\sum_{\lambda\in\sigma(K_\omega)} \lambda^n
(1+\alpha\tau\lambda^{-1})^{-1}|g_\lambda\rangle\langle
f_\lambda|\,.
\end{equation}

For the particular case $\omega=0$, $1+\alpha\tau>0$, we conclude
that
\begin{eqnarray}\label{A.19}
&&\hspace{-40pt}\lim_{n\to\infty} (K_0)^n u(t)= g_1(t)\langle
f_1|u\rangle
\nonumber\\
&&\hspace{25pt}=(1+\alpha\tau)^{-1}\big(u(\tau)+\alpha\int^\tau_0
\mathrm{d}s u(s)\big)\,,
\end{eqnarray}
as claimed in (\ref{1.7}).

To find the rate of convergence, under the same conditions, we
subtract $P_1$ and use
\begin{equation}\label{A.20}
\| K_0^n-P_1\|\leq \sum_{\lambda\in\tilde{\sigma}(K)} |\lambda|^n
|(1+\alpha\tau\lambda^{-1})|\, \|g_\lambda\|\, \|f_\lambda\|\,.
\end{equation}
Computing the norms in (\ref{A.20}) one uses
\begin{equation}\label{A.21}
|\lambda+\alpha\tau|^{-1}|\lambda|\, \|g_\lambda\|\, \|f_\lambda\|=
|\lambda+\alpha\tau|^{-1}|\lambda|\big(1+|\lambda+\lambda^\ast
-2|\lambda|^2|^{-1}(1-|\lambda|^2)\big)\leq (\alpha\tau)^{-1}
\end{equation}
for all $\lambda\in\tilde{\sigma}(K)$. The last inequality can be
checked directly using that $\lambda = \mathrm{e}^z$ satisfies
(\ref{2.18}). Hence
\begin{equation}\label{A.22}
\|K_0^n-P_1\|\leq\sum_{\lambda\in\tilde{\sigma}(K)} |\lambda|^n
(\alpha\tau)^{-1}\cong (\lambda_\mathrm{max})^n (\alpha\tau)^{-1}\,.
\end{equation}
In particular,
\begin{equation}\label{A.23}
\int^\tau_0 \mathrm{d}s\big(K_0^n
u(s)-\bar{u}\big)^2\leq(\lambda_\mathrm{max})^{2n}(\alpha\tau)^{-2}
\int^\tau_0 \mathrm{d}s u(s)^2\,,
\end{equation}
which is a more explicit bound than (\ref{A.6}).

\end{appendix}

\end{document}